\documentclass[12pt,preprint]{aastex}

\shorttitle{Cold Disks}
\shortauthors{Brown et al.}

\begin{document}


\title{Cold Disks: Spitzer Spectroscopy of Disks around Young
Stars with Large Gaps}


\author{{J.M. Brown\altaffilmark{1}, G.A. Blake\altaffilmark{2},
C.P. Dullemond\altaffilmark{3}, B. Mer{\'{\i}}n\altaffilmark{4},
J.C. Augereau\altaffilmark{4}, A.C.A. Boogert\altaffilmark{6},
N.J. Evans, II\altaffilmark{7}, V.C. Geers\altaffilmark{4},
F. Lahuis\altaffilmark{4,8}, J.E. Kessler-Silacci\altaffilmark{7},
K.M. Pontoppidan\altaffilmark{2}, E.F. van Dishoeck\altaffilmark{4}}}
\altaffiltext{1}{Division of PMA, MC 105-24, California
Institute of Technology, Pasadena, CA 91125, jmb@astro.caltech.edu}
\altaffiltext{2}{Division of GPS, MC 170-25, California
Institute of Technology, Pasadena, CA 91125}
\altaffiltext{3}{Max-Planck-Institut fur Astrophysik, Koenigstuhl 17,
69117 Heidelberg, Germany}
\altaffiltext{4}{{Leiden Observatory, Leiden University, PO Box 9513, 2300 RA Leiden, The Netherlands}}
\altaffiltext{5}{Laboratoire d'Astrophysique de l'Observatoire de Grenoble, BP 53, 38041 Grenoble Cedex 9, France}
\altaffiltext{6}{{Caltech/IPAC, MC 100-22, Pasadena, CA 91125}}
\altaffiltext{7}{Department of Astronomy, University of Texas at Austin, 1
University Station C1400, Austin, TX 78712-0259}
\altaffiltext{8}{SRON National Institute for Space Research, PO Box 800, 9700 AV
Groningen, The Netherlands }



\begin{abstract}

We have identified four circumstellar disks with a deficit of dust
emission from their inner 15-50 AU. All four stars have F-G spectral type, 
and were uncovered as part of the Spitzer Space Telescope ``Cores to Disks''
Legacy Program Infrared Spectrograph (IRS) first look
survey of $\sim$100 pre-main sequence stars. Modeling of the
spectral energy distributions indicates a reduction in dust density by
factors of 100-1000 from disk radii between $\sim$0.4 and 15-50
AU, but with massive gas-rich disks at larger radii. This large contrast
between the inner and outer disk has led us to use the term `cold
disks' to distinguish these unusual systems. However, hot dust
[0.02-0.2 $M_{\rm moon}$] is still present close to the central star
($R \leq$0.8 AU). We introduce the 30/13 \micron\, flux density ratio
as a new diagnostic for identifying cold disks. The mechanisms for
dust clearing over such large gaps are discussed. Though rare, cold
disks are likely in transition from an optically thick to an optically
thin state, and so offer excellent laboratories for the study of
planet formation.

\end{abstract}




\section{Introduction}

The evolutionary processes transforming massive, gas-rich
circumstellar disks into tenuous, gas-poor debris disks are still not
well understood. During this crucial interval, planets form and the
remaining disk material is accreted or dispersed. The path by which
this transition proceeds remains uncertain. Do disks clear uniformly
throughout via grain growth and settling or is the process accelerated
in the inner regions forming gaps (e.g. \citealt{alexander06})?
Evidence of dust clearing should be visible in the infrared (IR)
spectral energy distribution (SED).  The {\it Spitzer Space
  Telescope}, with its wide wavelength coverage and increased
sensitivity, is starting to reveal a new population of disks with
unusual SEDs.

Lack of mid-IR excess emission from disks has been interpreted as a
sign of dust clearing since the first disk SEDs were observed with
IRAS \citep{strom89}. Further examples and some spectra (e.g. HD
100546, \citealt{bouwman03}) were obtained of a few Herbig Ae/Be stars
with the ISO satellite. In this paper, we designate disks with this
characteristic missing mid-IR emission as `cold disks' due to the lack
of emission from warm dust. These SEDs morphologically fall between
the SEDs of classical optically thick disks, which show excess
emission throughout the IR, and debris disks characterized by very
weak far-IR excesses, leading them to be considered as transitional
objects between the two classes. 

Spitzer spectra are particularly important to characterize the sharp
rise, since they cover the critical 8-30 \micron\, range. The earliest
example discovered with Spitzer was CoKu Tau 4 modelled to have an
inner hole within 10 AU (\citealt{forrest04},
\citealt{dalessio05}). Three additional T Tauri stars, TW Hya, GM Aur
and DM Tau \citep{calvet02,calvet05}, have been identified as also
containing cleared inner regions. One of the most exciting proposed
explanations for these cold disks is that a planet has cleared a gap
in the disk and thus that these inner holes may trace the presence of
planetary systems \citep{varniere06}.

In this paper we present Spitzer spectra of four young stars of F-G
spectral type that show a deficit of dust emission from the inner
15-50 AU of the disk yet strong excesses longward of 20 \micron
. These cold disk sources were identified in the first look survey of
the Cores to Disks (c2d) Spitzer Legacy Project \citep{evans03},
comprising $\sim$100 young ($<$ 10 Myr) stars, mainly T Tauri K
and M stars, with circumstellar disks. All four sources are
characterized by a steep, $\sim$10$\times$ rise in flux beginning at
15 \micron, indicating a sudden change in disk properties at a
specific radius in a manner similar to that seen in the T Tauri
transitional disks. Such objects are rare but provide an important
window into disk evolution and planet formation processes.

\section{Observations}

The 5-35 \micron\, spectra were taken with the Infrared Spectrometer
(IRS; \citealt{houck04}) on Spitzer using both low (R=160,
$\lambda_{SL}$:5.2-14.5$\mu$m, $\lambda_{LL}$:14.0-38.0$\mu$m) and
high (R=600, $\lambda_{SH}$:9.9-19.6$\mu$m,
$\lambda_{LH}$:18.7-37.2$\mu$m) resolution modules. Spectra were
extracted from the Spitzer Science Center (SSC) Basic Calibrated Data
(BCD) images, generated by pipeline S13. For the low resolution
spectra, the SSC pipeline full aperture extraction was used. For the
high resolution modules, the c2d extraction, based on a combined
sinc fitting of the spectral trace to account for bad pixels and
background emission, was used (see \citealt{lahuis06} for further
details\footnote{Lahuis et al. 2006 is available at:
  http://ssc.spitzer.caltech.edu/legacy/c2dhistory.html}).

MIPS SED spectra, taken as part of the c2d project, are included in
the SEDs for all four cold disk sources. For each source the BCDs were
coadded using MOPEX.  The coadded images were extracted with IRAF,
using an optimized extraction with a 3 or 5 column wide aperture.

\section{Distinguishing characteristics}

The large ($\sim$100 object) c2d IRS first look sample allows
comparison between the four cold disk sources and the remaining
systems to identify trends which might be diagnostic of their
evolutionary state. The small number of cold disk sources ($<$ 5\%
identified), both in the c2d sample and in the literature, suggests
that this condition is rare either due to rapid evolutionary
timescales undergone by all stars or due to an unusual condition
unique to a small sample of stars.

\begin{figure}
\vspace{-3cm}
\includegraphics[scale=.80, angle=90]{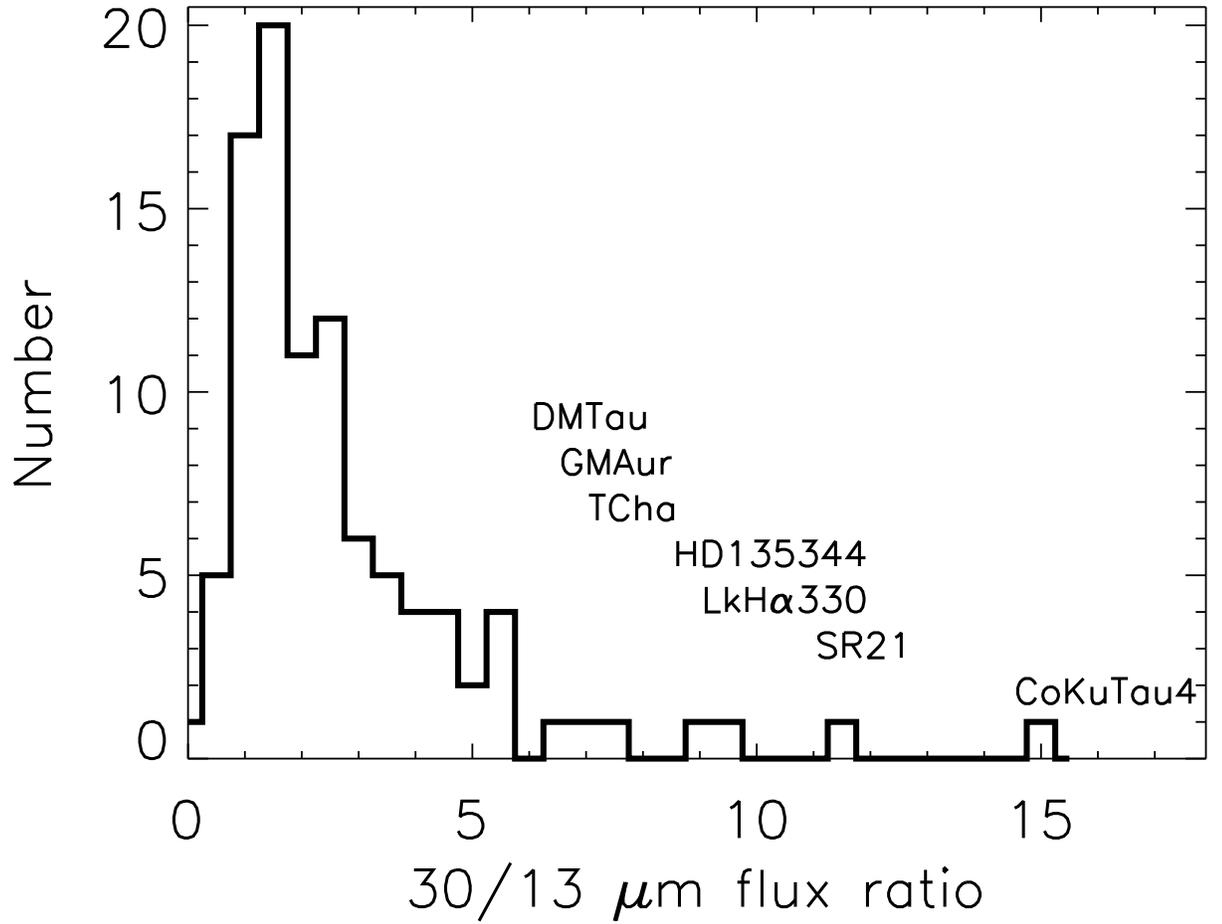}
\caption{Spitzer IRS 30/13 $\mu$m flux ratios for the c2d first look
disk sample, along with those of DM Tau, GM Aur and CoKu Tau 4.
The cold disks have much larger 30/13 \micron\, ratios than
does the majority of the sample. Outliers are labeled. \label{fig:1330}}
\end{figure}

The four cold disk sources are clearly differentiated from the
majority of the c2d star+disk systems by the $\sim$10 fold increase in IR
flux between 10 and 30 \micron. To characterize this increase in dust
emission, we have used the 30/13 \micron\, flux ratio (see Figure
\ref{fig:1330}). These wavelengths have been chosen to avoid strong
silicate features, but include the full increase in continuum
emission. The majority of the sample has emission that increases by a
factor of 2.3 $\pm$ 1.4 between 13 and 30 \micron\, while the
transitional disks rise by factors of 5-15.

Interestingly, out of a sample comprised of predominantly low-mass
stars, the cold disk sources are all of intermediate mass with
spectral types of F and G. The majority of K and M stars and the
higher mass A and B stars have low 30/13 \micron\, ratios. However,
there are only a handful of A and B stars in our sample so it is
difficult to draw any significant conclusions about such stars.

All four cold disk sources show polycyclic aromatic hydrocarbon (PAH)
features, particularly at 11.3 \micron. Such emission is uncommon in
the c2d sample with only $\sim$10\% displaying PAH features
(\citealt{geers06}, see their Figure 5 for blow-ups of the PAH
bands). The presence of PAH features may be enhanced in
these cold disks because the lack of dust in the inner disk lowers the
mid-IR continuum flux creating a stronger line-to-continuum ratio,
thus facilitating detection.

The four disks are also characterized by weak to non-existent 10
\micron\, amorphous silicate features. LkH$\alpha$ 330 is the only one
that shows an unambiguous, though low contrast, 10
\micron\, feature. The spectrum of HD 135344 includes the wavelength
region of the full 10 \micron\, band but shows no silicate
feature. The spectra of SR 21 and T Cha begin at 10 $\mu$m, but appear
to have only weak silicate emission, if any. The broader 20 \micron\,
silicate feature is harder to isolate from continuum dust emission,
particularly with the sharp rise in the SED beyond 15
\micron. However, some 20 \micron\, emission does seem to be
present. This feature traces colder regions of the disk than does the
10 \micron\, feature, and thus indicates the presence of amorphous
silicates further out in the disks.

The H$\alpha$ equivalent width is often used as a tracer of
accretion. Different dividing lines between non-accreting weak-line T
Tauri stars (wTTs) and classical T Tauri stars (cTTs) have been
proposed but it makes little difference to the classifications in
these cases. SR 21 (0.54 \AA\, in absorption, \citealt{martin98}) and
T Cha (2-10 \AA, \citealt{gregorio92}, \citealt{alcala95}) are
nominally wTTs, although T Cha is highly variable and close to the
cutoff.  LkH$\alpha$ 330 (11-20 \AA, \citealt{fe95}, \citealt{ck79})
and HD 135344 (17.4 \AA, \citealt{acke05}) are clearly cTTs.

\section{Modeling}

Modeling is necessary in order to interpret the SEDs in terms of the
physical structure of the disk. In particular, can a disk model with
a gap accurately reproduce the SEDs, including the steep rise between
13 and 30 \micron ? Is such a fit possible without resorting to a gap?

The disks were modeled with the 2-D radiative transfer code RADMC
\citep{dd04}. The steep factor of 10 rise in flux between 13 and 30
\micron\, prevents these disks from being fit well by conventional
disk models. We introduce a very wide gap with an inner radius,
$R_{\rm Gap,in}$, near 1 AU and an outer radius, $R_{\rm Gap,out}$,
near 30 AU (see Table \ref{table:model} for specific values). In order
to model the steep change in emission associated with $R_{\rm
  Gap,out}$, grid refinement is introduced at $R_{\rm Gap,out}$ as
well as the inner dust rim, $R_{\rm Disk,in}$.  A physical reduction
in dust density is only one possible scenario which could result in
this SED shape (see \S 5).

Input parameters for these models include the stellar mass, $M_*$,
radius, $R_*$, and effective temperature, $T_{\mathrm{eff}}$.  Kurucz
models are used for the stellar photospheres. Where possible, values
for stellar and disk properties are taken from the literature (see
Table \ref{table:model}). The \citet{siess00} pre-main sequence
stellar tracks were used to check that the parameters were
consistent. The effects of modest differences in the stellar
properties on the mid-IR portions of the spectra are small. For
example, a 100 K change in the stellar temperature results in a change
in $R_{\rm Gap,out}$ of 2-3 AU. Optical and (sub)millimeter photometry
is often not simultaneous and is dereddened using the extinction law
of \citet{draine03}.

The disks are assumed to be flared with surface height $H$ such that
$H/R \propto R^{2/7}$ as in \citet{cg97}. The models do not calculate
the hydrostatic equilibrium self-consistently and here the pressure
scale height is set at the outer disk edge, $R_{\mathrm{disk}}$, which
is assumed to be 300 AU for all the models. H/$R_{\mathrm{disk}}$
values slightly lower than predicted by hydrostatic equilibrium are
needed to fit the outer disks of both HD 135344 and T Cha, which
produces a smaller silicate feature even in the models with no
gap. This could indicate dust settling in the outer disk, which might
be expected from the older ages (8-10 Myr vs 2-3 Myr) infered by the
\citet{siess00} tracks. The dust composition is set in all models to a
silicate:carbon ratio of 4:1. The grain sizes range from 0.01 to 10
$\mu$m with a power-law index of -3.5 and a dust mass of $M_{\rm
  Dust,small}$. The upper size limit of 10 \micron\, is large compared
to interstellar grains \citep{mathis77}, but was necessary to account
for the lack of any 10 \micron\, amorphous silicate feature.  This
indicates a skewing of the size distribution likely through grain
growth. No crystalline silicates are included. A midplane layer of 2
mm grains containing 90\%~ of the total dust mass, with the remaining
10\%~ in $M_{\rm Dust,small}$, has been added to account for the
(sub)millimeter slope (as in \citealt{dd04}). The inner edge of the
disk, $R_{\rm Disk,in}$, was set at the radius where the dust
sublimation temperature of $T_{\mathrm{dust}}$ $\sim$ 1500 K is
reached. A slightly puffed up inner rim was included with $H/R_{\rm
  Disk,in}$ being set at 0.03, just above the hydrostatic equilibrium
value. The disk inclinations are not well constrained so
$i \sim$ 30$^\circ $ was used arbitrarily for SR 21 and LkH$\alpha$
330. T Cha is more edge on so $i \sim$ 75$^\circ $ was adopted
\citep{alcala93} and HD 135344 is more face-on with $i \sim$ 10$^\circ
$ \citep{dent05}. However, until the disk is nearly edge on and begins
to obscure the central star, varying the inclination produces little
effect on the gap properties.

The optimal size of the gap necessary to fit the mid-IR is found by
minimizing the $\chi ^2$ of the fit (see Figure
\ref{fig:chi}). $R_{\rm Gap,out}$ is tightly constrained by the steep
rise at 13 \micron, while $R_{\rm Gap,in}$ is less well constrained. The
dust density in the gap was reduced by 10$^{-6}$ compared to the
equivalent models with no gaps, and the minimum reductions needed were
factors of 10-100.

All of our sources have 1-10 \micron\, excesses which require
0.02-0.24 M$_{\rm moon}$ of dust between $R_{\rm Disk,in}$ and $R_{\rm
  Gap,in}$, a fraction of only 10$^{-7}$ to 10$^{-6}$ of the total
refractory dust mass available. This matter is optically thick to the
stellar light at the midplane. If we remove {\em all} matter from the
disk inward of $R_{\rm Gap,out}$, then the near-IR fluxes are
underpredicted. These excesses clearly show that at least some dust
must be close to the star. We model the 1-10 $\mu$m excess by keeping
the inner disk intact and only inserting a gap from
$R_{\mathrm{Gap,in}}$ to $R_{\mathrm{Gap,out}}$, where
$R_{\mathrm{Gap,in}} > R_{\mathrm{Disk,in}}$. For T Cha, this results
in a particularly small, hot dust region inside the gap which would
have a very short lifetime, of order thousands of years at most,
without a continual influx of material.

\section{Discussion}

We have identified four cold disks around F and G type stars with
unusually steep flux rises between 10 and 30 \micron , whose SEDs can
only be modelled with wide gaps of inner radii of 0.2-0.8 AU and outer
radii 15-50 AU. These gaps are generally larger than those inferred
for the 4 T Tauri stars with similar 30/13 \micron\, ratios found so
far, which have outer gap radii of 10-24 AU. Another difference is
that our sources have small 1-10 \micron\, excesses which demand that
hot dust exists between the inner edge of the gap and the star,
i.e. our sources have gaps rather than holes. The statistics on
sources of different spectral types are still too small to conclude
whether this is a general trend or peculiar to our sources, and
whether this extends to Herbig Ae/Be stars.

Dust clearing related to planet formation would be one of the most
exciting explanations for the observed SEDs. There is also the
possibility of the disk being disrupted and the dust cleared by a
stellar companion. HD 135344 and SR 21 are part of wide binaries with
separations of 20\farcs4 and 6\farcs4, respectively
(\citealt{coulson95}; \citealt{pgs03}). Although no companions are
currently known within the modeled $R_{\mathrm{disk}}$, close binaries
($<$50 AU) cannot be ruled out for any of the sources.

Another proposed scenario for quickly clearing the inner disk region
is photoevaporation (\citealt{clarke01}, \citealt{alexander06}). This
physical process occurs when the photoevaporation rate driven by the
ionizing flux from the central star matches the viscous accretion
rate, resulting in an inner hole. The predicted size of the inner hole
is given by $R_{\rm g}=GM_*/c_{\rm s}^2$, with $T \sim$ 10$^4$ K to
give $c_s \sim $ 10 km/s, leads to predicted hole radii of 13-18 AU for
1.5-2 M$_\odot$ stars, although a more rigorous examination of the gas
dynamics revealed that this equation over-estimates the inner hole
size \citep{liffman03}. However, models of Herbig Ae stars (M$_*$=2.5
M$_\odot$) that combine photoevaporation with viscous evolution and
differential radial motions of dust and gas predict that the inner
disk clears quickly but leaves gas-poor dust rings at 10-100 AU
\citep{takeuchi05}. The accretion rates for all four cold disks, with the
possible exception of SR 21, are too high to make this scenario
likely.

An alternative to physically removing the dust is to let it grow
beyond the size at which it efficiently radiates as a blackbody so
that it no longer emits strongly in the mid-IR \citep{tanaka05}. There
is general evidence for grain growth in disks from both mid-IR and
millimeter data (\citealt{kessler-silacci06},
\citealt{rodmann06}). The lack of strong 10 \micron\, amorphous
silicate features also points to grains having
grown beyond interstellar sizes. There has been much recent modeling
work on clearing disks through grain growth. \citet{dd05} found that
cold disk SEDs could be produced by dust coagulation but the
timescales were too fast. Replenishment of the dust was necessary to
match observed lifetimes of the disks. If replenishment processes such
as fragmentation occur preferentially in the innermost region due to
higher temperatures and densities \citep{kb04}, this might be
sufficient to produce the fraction of a lunar mass needed close to the
star. \citet{rice06} invoke dust filtration by an embedded planet
whereby large grains pile up at the outer edge of the gap while small
grains and gas pass through, thus accounting for both the hot small
dust grains needed to fit the near-IR and the high mass accretion
rates. The high spatial resolution of ALMA should be able to test
these scenarios by searching for and imaging the emission from
(sub)millimeter to centimeter sized grains.

\acknowledgments{Support for this work, part of the Spitzer Legacy Science
  Program, was provided by NASA through contracts 1224608, 1230779 and
  1256316 issued by the Jet Propulsion Laboratory, California
  Institute of Technology, under NASA contract 1407. Astrochemistry in
  Leiden is supported by a NWO Spinoza grant and a NOVA grant, and by
  the EU RTN-PLANETS (HPRN-CT-2002-00308). B.M. acknowledges the
  financial support from the Fundacion Ramon Areces (Spain). We thank
  the Lorentz Center in Leiden for hosting several meetings that
  contributed to this paper.}

\bibliographystyle{apj}

\clearpage

\begin{deluxetable}{lcccccccccccc}
\tablecolumns{13}
\setlength{\tabcolsep}{0.03in}
\tablewidth{0pt}
\tabletypesize{\footnotesize}
\tablecaption{\label{table:model} Model parameters}
\tablehead{\colhead{Source} & \colhead{Spectral} & \colhead{$A_v$} &
\colhead{Distance}  & \colhead{$M_{\mathrm{*}}$} &
\colhead{$T_{\mathrm{eff}}$} & \colhead{$L_{\mathrm{*}}$} &
\colhead{$R_{\mathrm{Disk,in}}$} &
\colhead{$R_{\mathrm{Gap,in}}$} &
\colhead{$R_{\mathrm{Gap,out}}$} & \colhead{$M_{\rm Dust,small}$} &
\colhead{$M_{\rm inner}$} & \colhead{$H_{\mathrm
p}(R_{\mathrm{disk}})/$}  \\
& \colhead{Type} & (mag) &\colhead{(pc)}& \colhead{($M_{\odot}$)} & \colhead{(K)} &
\colhead{($L_{\odot}$)} & \colhead{(AU)} & \colhead{(AU)} &
\colhead{(AU)} & \colhead{(10$^{-6}$ $M_{\odot}$)} & \colhead{($M_{\rm lunar}$)}
& \colhead{$R_{\mathrm{disk}}$}}
\startdata
LkH$\alpha$ 330 & G3$^a$ & 1.8$^a$ & 250$^b$ & 2.5$^c$ & 5800 & 16 & 0.27
& 0.8 & 50 & 5 & 0.24 & 0.15\\
SR 21 & G2.5$^d$ & 9$^d$ & 160 & 2.5$^d$ & 5800 & 24 & 0.25 & 0.45 & 18 &
15 & 0.10 & 0.17\\
HD 135344 & F4$^e$ & 0.5$^f$ & 84$^e$ & 1.8 & 6600$^e$ & 6.8 & 0.18 & 0.45 &
45 & 5 & 0.10 & 0.13\\
T Cha & G8$^g$ & 1.5$^g$ & 66$^h$ & 1.5$^g$ & 5600 & 1.4 & 0.08 & 0.2 &
15 & 3 & 0.025 & 0.11
\enddata
\tablenotetext{1}{$a$ - \citet{ck79}, $b$ - \citet{enoch06}, $c$ -
  \citet{ob95}, $d$ - \citet{pgs03}, $e$ - \citet{dbr97}, $f$ -
  \citet{malfait98}, $g$ - \citet{alcala93}, $h$ - \citet{wichmann98}}
\end{deluxetable}

\clearpage



\clearpage
\begin{figure}
\vspace{-3cm}
\plotone{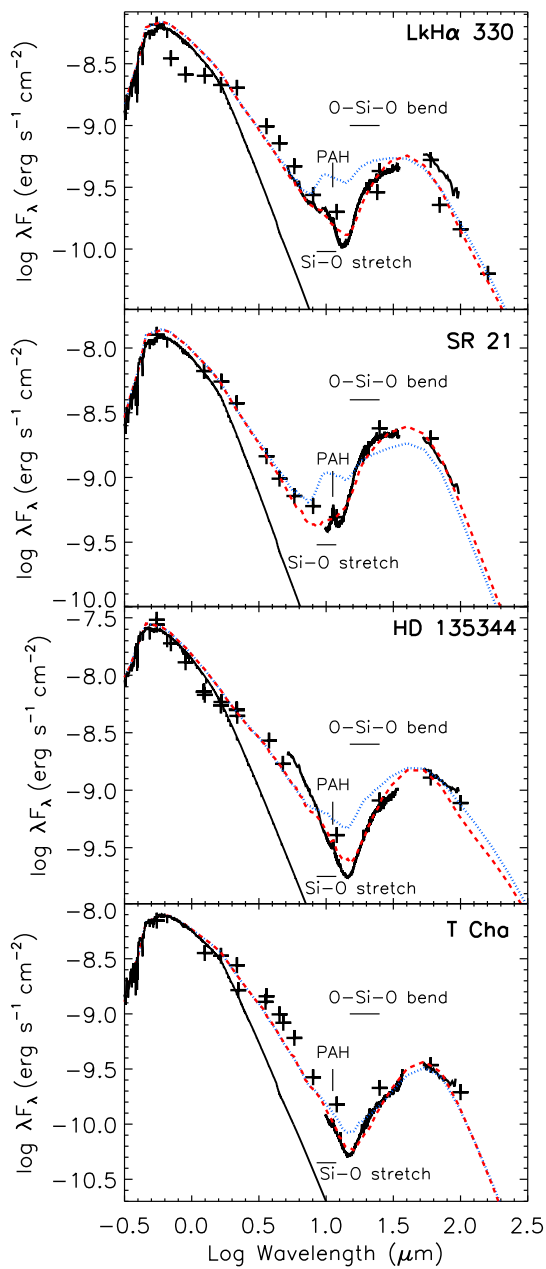}
\caption{Spitzer IRS and MIPS spectra of the four cold disk sources
reported here. The dashed line is the best fit star+disk model with a
gap (see Table 1) and the dotted line is the equivalent model with no
gap. JHK photometry is from 2MASS. IRAS photometry and our Spitzer
photometry is shown in the mid- to far-IR. Optical and submillimeter
photometry is from the literature (\citealt{alcala93}; \citealt{am94};
\citealt{fe95}; \citealt{henning93}; \citealt{ob95}).
\label{fig:spectra}}
\end{figure}

\clearpage

\begin{figure}
\plotone{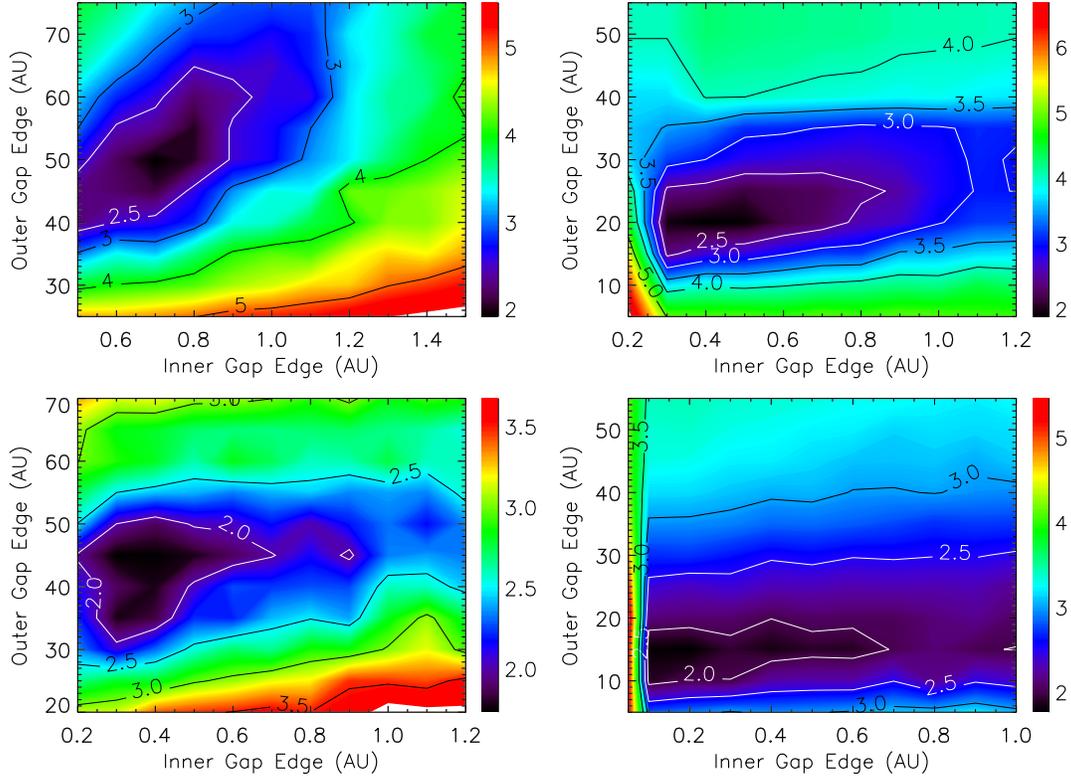}
\caption{Chi squared contour maps of 2-D disk model fits to the SEDs
with different inner and outer gap radii (clockwise from top left:
LkH$\alpha$ 330, SR 21, T Cha and HD 135344).
\label{fig:chi}}
\end{figure}


\end{document}